\documentclass[aps,prb,twocolumn,superscriptaddress]{revtex4}
\usepackage{amssymb}
\usepackage{graphicx}
\usepackage{amsmath}
\usepackage{bm}

\begin{document}

\title{Machine Learning of Frustrated Classical Spin Models. II. Kernel Principal Component Analysis}

\author{Ce Wang}
\affiliation{Institute for Advanced Study, Tsinghua University, Beijing, 100084, China}
\author{Hui Zhai}
\affiliation{Institute for Advanced Study, Tsinghua University, Beijing, 100084, China}
\affiliation{Collaborative Innovation Center of Quantum Matter, Beijing, 100084, China}

\date{\today }

\begin{abstract}

In this work we apply the principal component analysis (PCA) method with kernel trick to study classification of phases and phase transition in classical $XY$ models in frustrated lattices. Comparing to our previous work with linear PCA method, the kernel PCA can capture non-linear function. In this case, the $Z_2$ chiral order of classical spins in these lattices are indeed a non-linear function of the input spin configurations. In addition to the principal component revealed by linear PCA, the kernel PCA can find out two more principal components using data generated by Monte Carlo simulation at various temperatures at input. One of them relates to the strength of the $U(1)$ order parameter and the other directly manifests the chiral order parameter that characterizes the $Z_2$ symmetry breaking. For a temperature resolved study, the temperature dependence of the principal eigenvalue associated with the $Z_2$ symmetry breaking clearly shows a second order phase transition behavior.

\end{abstract}

\maketitle

\section{Introduction}

Recently there is an increasing interests in applying machine learning to classify phase of matter\cite{linearPCA,Ising_PCA,Ising_nn,Confusion,Ising_BM,Ising_XY_VAE,Ising_SVM,SqXY_PCA,Ising_SU(2)_nn,CNN_Fermions1,CNN_Fermions2,CNN_uslearning,QLTML,windingnumberML,Fermion_PCA,Zhangyi_1,Hubbard_uslearning,KT_nn}, including both quantum and classical phases with different order parameters\cite{linearPCA,Ising_PCA, Ising_nn,Confusion,Ising_BM,Ising_XY_VAE,Ising_SVM,SqXY_PCA,Ising_SU(2)_nn,CNN_Fermions1,CNN_Fermions2,CNN_uslearning,Fermion_PCA,Hubbard_uslearning,KT_nn}, as well as topological phases\cite{QLTML,windingnumberML,Zhangyi_1}. The schemes involved include both supervised learning \cite{Ising_nn, Confusion, Ising_SVM,Ising_SU(2)_nn,CNN_Fermions1,CNN_Fermions2,QLTML,windingnumberML,Zhangyi_1} and unsupervised learning\cite{linearPCA,Ising_PCA,Ising_BM,Ising_XY_VAE,CNN_uslearning,Fermion_PCA,Hubbard_uslearning,SqXY_PCA,KT_nn}. For example, previous works have applied unsupervised learning technique such as the principal component analysis (PCA) to analyze data generalized by classical Monte Carlo simulation for both Ising and $XY$ model at different temperatures\cite{linearPCA,Ising_PCA,SqXY_PCA}, and they have shown that the PCA method can clearly distinguish the difference between a high temperature disordered phase and the low-temperature ordered phase. 

In the first paper of this series\cite{linearPCA}, we have applied the PCA method to frustrated classical spin models\cite{FXY_theory0,FXY_theory1,FXY_theory2,FXY_theory3,FXY_theory4,FXY_theory5,FXY_theory6},such as $XY$ model in a triangular and union jack lattices\cite{Monte_tri,Monte_tri_uj}. Unlike the $XY$ model in the square lattice\cite{XY_Tc,XY1,XY2}, the low temperature phase in these frustrated models displays both $U(1)$ and $Z_2$ order parameters, and as temperature increases, there exhibit two phase transitions across which the $U(1)$ and $Z_2$ order parameters disappear separately. For instance, for the union jack lattice, the low temperature phase contains both two orders, and the intermediate phase only has $U(1)$ order and the $Z_2$ order vanishes, and the high temperature phase has no orders. We show that the PCA method applied to data generated at all temperatures can clearly distinguish all these three phases. By applying PCA to temperature resolved data, the principal values as a function of temperature can also reveal the location and the nature of the phase transitions. More importantly, we also design a simple toy model to understand why the PCA method can work in such application of recognizing phase of matter. 

Nevertheless, at the end of this paper, we also mentioned the limitation of the PCA method. One limitation is that it does not allow a direct readout of the order parameter. However, we should remark that in some cases, the PCA method does allow a direct readout of the order parameter, and the simplest example is the Ising model. In this case, the input data is a set of spin configurations $x_n$ as $\{s_i,\{i=1,\dots,L\}\}$ where $s_i$ is the Ising spin at each site and takes either $+1$ or $-1$. PCA will find out one principle component with eigenvector $(1/L)\{1,1,\dots,1\}$. The projection of each $x_n$ into this principle component is simply $(1/L)\sum_{i}s_i$, and this is nothing but the total magnetization and is the order parameter for the Ising model, which captures the $Z_2$ symmetry breaking. 

\begin{figure}[t]
\includegraphics[width=3.4 in]
{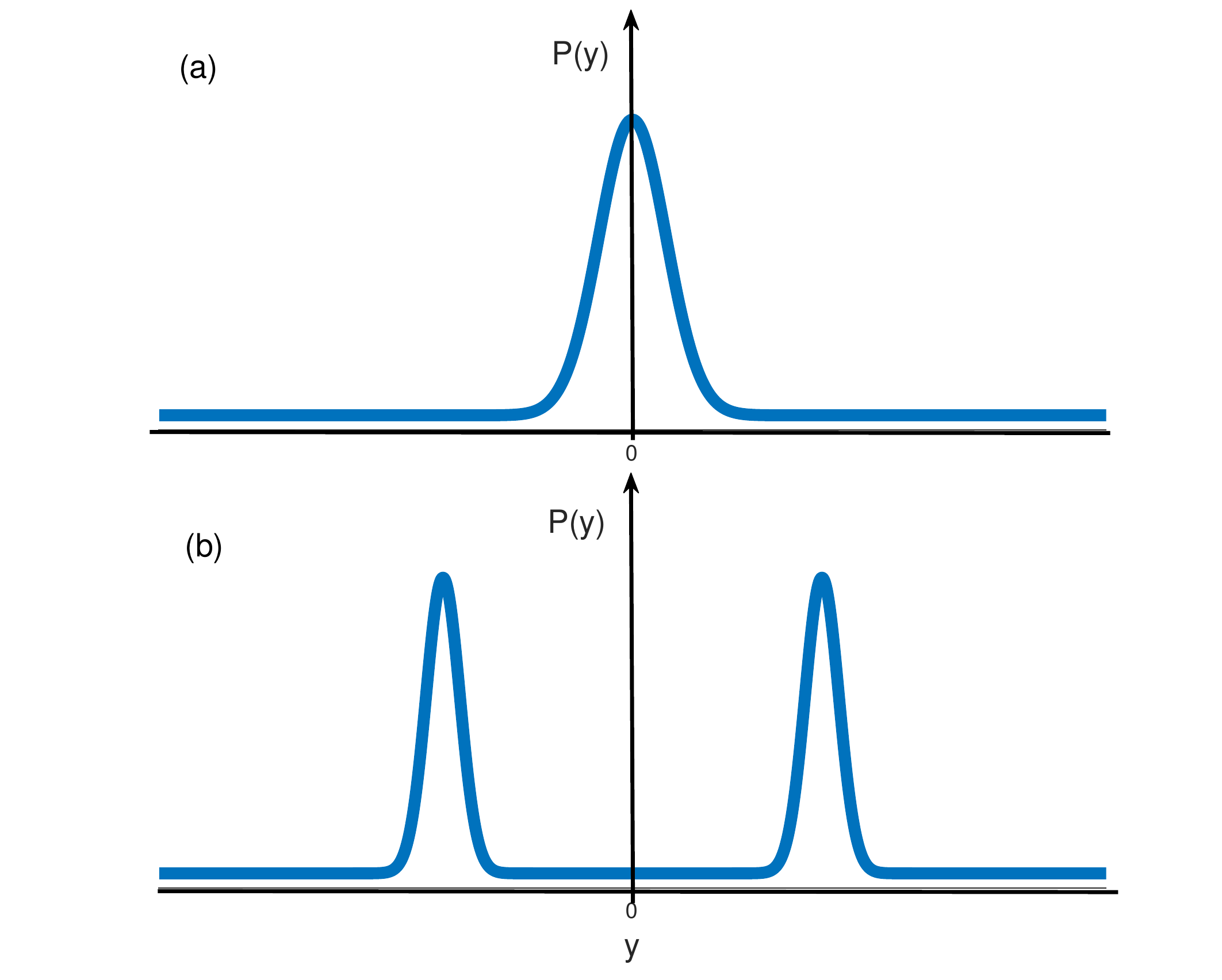}
\caption{Schematic of mathematical description of $Z_2$ symmetry breaking: Histogram $P(y)$ for (a) high temperature data set without a $Z_{2}$ symmetry breaking and (b) low temperature data with a $Z_{2}$ symmetry breaking. }
\label{z2}
\end{figure}

Taking this $Z_2$ symmetry breaking as an example, let us be more concrete about a mathematical description of the symmetry breaking for a general situation. Considering a function $f(x)$ with a spin configuration $x$ as the input and a single number $y=f(x)$ as the output, and for a set of data set $\{x_n\}$, we consider the probability $P(f(x)=y)$ with $x$ running over all data within the data set. $P(y)$ is also known as the histogram. If $\{x_n\}$ is a high temperature data set, $P(y)$ will behave like Fig.\ref{z2} which is a symmetric function peaked at $y=0$; however, for low temperature data set $\{x_n\}$, $P(y)$ should be a double peaked symmetric function as shown in Fig.\ref{z2}. In the case of Ising model, $f(x)$ can be simply chosen as averaged total magnetization $(1/L)\sum_{i}s_i$.

However, in the case of the triangular and union jack lattices, the $Z_2$ degree of freedom is the chirality order, which describes clockwise or anti-clockwise spin rotation around each triangle. The PCA method can distinguish phases with different chirality, but it fails to directly characterize the $Z_2$ symmetry breaking as mentioned above, simply because the chirality order of this case is a non-linear function of the input, and consequently $f$ should also be a non-linear function. 

In machine learning, a so-called kernel trick is often used to capture non-linear characteristics of the order parameter. In this paper, we will apply the kernel PCA method to $XY$ model in different lattices, and we will show that the kernel PCA method can capture the strength of the $U(1)$ order and allow a direct readout of the $Z_2$ chirality order. We also design a similar toy model to understand how the kernel PCA method works here. 

\section{Model and Method}

Here we will first briefly review the model we consider and the kernel PCA method\cite{PRML}. The Hamiltonian of the classical XY model is given by
\begin{equation}
\mathcal{H}=J\sum\limits_{\langle ij\rangle}\boldsymbol{s}_{i} \cdot \boldsymbol{s}_{j},
\end{equation}
where $\boldsymbol{s}_{i} = (\cos\theta_{i}, \sin\theta_{i})$ is a classical planar spin defined at each site and $\theta_{i}\in(0,2\pi]$, $\langle ij\rangle$ denotes all the nearest neighboring bonds. Since they are classical models, the data we use are generated by classical Monte Carlo simulation. Such an algorithm can produce equilibrium spin configurations at different temperatures, denoted by $\{x_n\}$ ($n=1,\dots,N$). Each $x_n$ is a vector of $2L$-dimension organized as  
\begin{equation}
x_n=(\cos\theta_1, \dots,\cos\theta_L,\sin\theta_1,\dots,\sin\theta_L),
\end{equation} 
where $L$ is the total number of lattice sites and $N$ is the total number of data set.  $\{x_n\}$ ($n=1,\dots,N$) will be the data and the only input that we feed to computer. 
In our analysis below, in most cases we will use data generated at different temperatures as the input. In some cases specified as temperature resolved analysis, we use data with a given temperature as the input. 

The kernel PCA method is first to map the original data $x$ of $2L$-dimension into a $M$-dimensional data $\phi(x)$ with $M>2L$ by a mapping $\phi(x)$. The PCA for this new data set $\{\phi(x_{n})\}$ is called the kernel PCA. We can define $\bar\phi = (1/N)\sum_{n=1}^{N}\phi(x_n)$ is the average of all the data, and organize all data into a $N\times M$-dimensional matrix $\mathcal{X}$, where $\mathcal{X}_{nm}$ denotes the $m$-th component of  $\phi(x_n) - \bar{\phi}$. 
Now we define a $N\times N$-dimensional matrix $\mathcal{K} = \mathcal{X}\mathcal{X}^T$. It is easy to show that the matrix $\mathcal{K}$ has the same set of eigenvalue as a $M\times M$ dimensional matrix $\mathcal{S} = \mathcal{X}^{T}\mathcal{X}$. The linear PCA used in previous paper can be viewed as a special case where we diagnoalize matrix $\mathcal{S}$ with $\phi(x)=x$.  It is straightforward to show that 
\begin{equation}
\mathcal{K} = K-\frac{1}{N} E_{N,N}K -\frac{1}{N} KE_{N,N} +\frac{1}{N^2}E_{N,N}KE_{N,N}, \label{pca1}
\end{equation}
with $E_{a,b}$ a $a\times b$ matrix whose elements all set to unity and $K_{ij}=\phi(x_{i})\phi(x_{j})^T$ is denoted by $k(x_{i},x_{j})$. Here $k(x,y)$ is called the kernel function. In practices, instead of designing the mapping function $\phi(x)$, we usually directly design the kernel function $k(x,y)$, and it can be proved that each choice of $k(x,y)$ corresponds to a choice of $\phi(x)$. 

Since the matrices $\mathcal{K}$ and $\mathcal{S}$ have the same set of eigenvalues, for instance, for the same eigenvalue $\lambda$, the corresponding eigenvector for the matrix $\mathcal{K}$ is denoted by ${\bf v}$ and the corresponding eigenvector for the matrix $\mathcal{S}$ is denoted by ${\bf u}$, it can be shown that ${\bf u}$ and ${\bf v}$ are related by  
\begin{equation}
{\bf u} = \frac{1}{(N\lambda)^{1/2}}\mathcal{X}^{T}{\bf v}. \label{pca2}
\end{equation}
Suppose ${\bf v}_{i}$ is the eigenvector of $\mathcal{K}$ corresponding to the $i$th largest eigenvalue, following Eq. \ref{pca2}, the projection of a general data $\phi(x)-\bar\phi$ onto ${\bf v}_i$ is 
\begin{equation}
\mathcal{P}_{i}(x) = \frac{1}{(N\lambda_{i})^{1/2}}\sum_{j=k}^{N} v_{ik} (\phi(x)-\bar\phi)(\phi(x_{k})-\bar\phi)^T, \label{projection}
\end{equation}
where $v_{ik}$ is the $k$th component of ${\bf v}_{i}$. A consequence of Eq.\ref{projection} is that 
\begin{equation}
\mathcal{P}_i(x_j) \propto v_{ij}. \label{projection2}
\end{equation}
Below we will apply this kernel PCA algorithm with a quadratic kernel $k(x,y)=(xy^T)^2$ to the $XY$ model in several different lattices, respectively. 

\begin{figure}[t]
\includegraphics[width=3.0 in]
{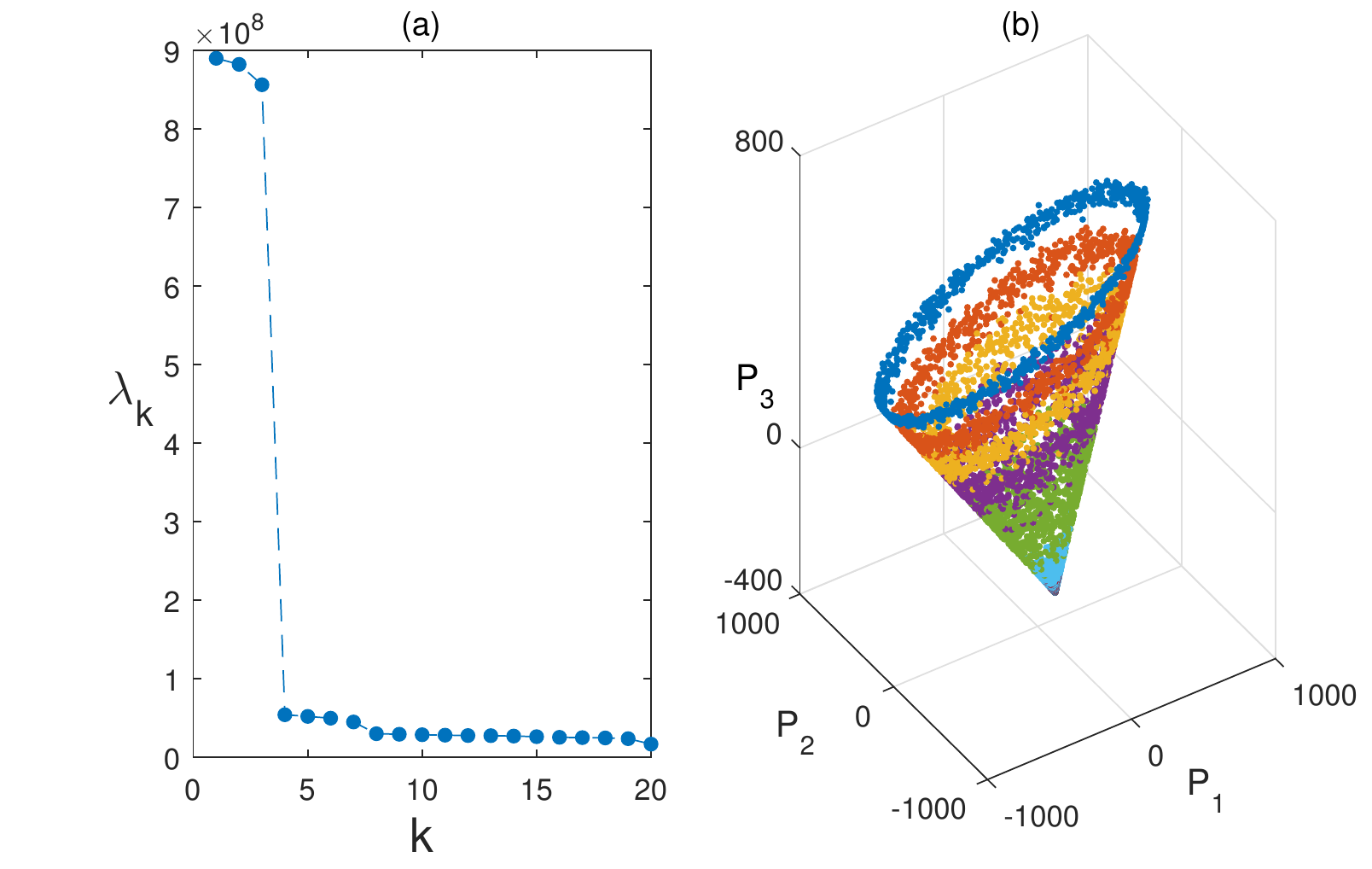}
\caption{(a)Eigenvalues of the kernel PCA applied to the $XY$ model on $L= 36\times36$ square lattice. The temperature of the data set ranges from $0.2J$ to $1.8J$, with $\Delta T = 0.2 J$. At each temperature, $1000$ samples are collected. There are three eigenvalues much larger than the others.  (b) The projection of the data into the subspace spanned by the first three largest eigenvectors. }
\label{sq}
\end{figure}

\section{Square Lattice Model}

We first apply this method to square lattice. The input data are generated at temperature ranged from well above transition temperature to well below transition temperature. The results from a quadratic kernel are shown in Fig. \ref{sq} (a), in which we find three eigenvalues that are much larger than all the rest. Their corresponding eigenvalues are denoted by ${\bf v}_1$ to ${\bf v}_3$, and the projection of the data $x_n$ into the subspace spanned by these three vectors forms a three dimensional vector $l_n=\{\mathcal{P}_1(x_n),\mathcal{P}_2(x_n),\mathcal{P}_3(x_n)\}$. These vectors are shown in Fig. \ref{sq}(b) and they look like a cone. 

In the previous work, we have developed a toy model to understand the outcome from the linear PCA. The toy model assumes that $p\%$ percent of data are completed ordered, that is to say, all spins at different sites are pointing to the direction $\theta$, and the other $(1-p)\%$ percent of data are completed disordered, that is to say, spins between any two sites are not correlated at all. Here we show that the same toy model can also be used to understand the meaning of this outcome from the kernel PCA. 

With this simplification of the input data, it is easy to show that the $\mathcal{K}$ matrix can be constructed as
\begin{equation}
\mathcal{K} = \begin{pmatrix}
K^{\text{low}}+c_{1}E_{pN,pN} & c_{2}E_{pN,(1-p)N}\\
c_{2}E_{(1-p)N,pN} & K^{\text{high}}+c_{3}E_{(1-p)N,(1-p)N}
\end{pmatrix}, \label{square_kpca}
\end{equation}
where 
\begin{align}
&K^{\text{low}}_{ij} =  \frac{L^{2}}{2}\left(1 + \cos\left(\frac{4\pi(i-j)}{pN}\right)\right)\\
&K^{\text{high}}_{ij}=L^{2}\delta_{ij}.
\end{align} 
and $c_1 = L^{2}(p^2-2p)/2$, $c_2 = -L^{2}p(1-p)/2$ and $c_3 =L^2 p^2/2$. Similarly, the eigenvectors corresponding to the largest two eigenvalues of $\mathcal{K}$ are of the form
\begin{align}
&{\bf v}_{1} \propto \left(\cos\left(\frac{4\pi}{pN}\right),...,\cos\left(\frac{4\pi n}{pN}\right),...,\cos(4 \pi ),0,...,0\right),\\
&{\bf v}_{2} \propto \left(\sin\left(\frac{4\pi}{pN}\right),...,\sin\left(\frac{4\pi n}{pN}\right),...,\sin(4 \pi ),0,...,0\right).
\end{align}
It can be shown that the corresponding projections onto ${\bf v}_1$ and ${\bf v}_2$ are
\begin{align}
&\mathcal{P}_{1}(x)  \propto\sum_{n,m=1}^{L}\cos( \theta_{m}+\theta_{n}), \\
&\mathcal{P}_{2}(x)  \propto\sum_{n,m=1}^{L}\sin( \theta_{m}+\theta_{n}). 
\end{align}
For fully correlated data, because $\theta_n=\theta_m=\theta$, therefore $\{\mathcal{P}_1(x),\mathcal{P}_2(x)\}\propto \{\cos(2\theta),\sin(2\theta)\}$, which forms a circle. For fully uncorrelated data, averaging over all sites gives rise to vanishing small numbers.  This part contains the information of $U(1)$ order parameter and it is pretty the same as the projection to the first two principal components in the linear PCA case.  

\begin{figure}[t]
\includegraphics[width=3.0 in]
{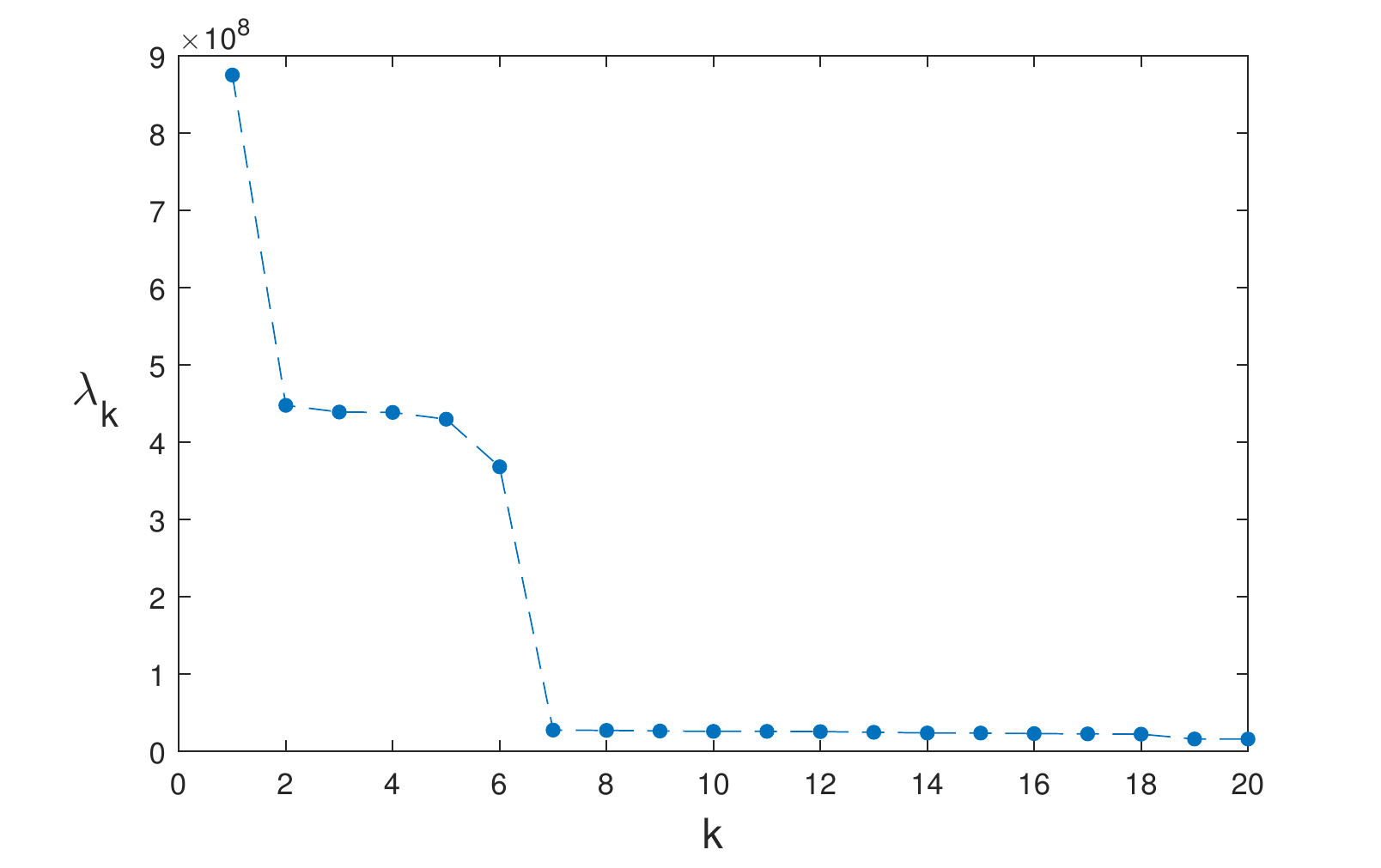}
\caption{The eigenvalues of kernel PCA for XY model on $L=36\times 36$ triangular lattice. The temperature of the data set ranges from $0.3J$ to $0.7J$, with $\Delta T = 0.05 J$. At each temperature, 1000 samples are collected.}
\label{tri_eigenvalue}
\end{figure}

Now we come to the projection onto ${\bf v}_3$. It can be shown that within this toy model 
\begin{equation}
{\bf v}_{3} \propto (1-p,1-p,....,1-p,-p,-p,...,-p). \label{square_v3}
\end{equation}
and therefore,
\begin{equation}
\mathcal{P}_{3}(x) \propto\sum_{n,m=1}^{L} \cos(\theta_{n}-\theta_{m})+\mathcal{C}  \end{equation}
where $\mathcal{C}$ is a constant. It is straightforward to show that 
\begin{align}
&\mathcal{P}_{1}(x) = \mathcal{A}^2 -\mathcal{B}^{2},\\
&\mathcal{P}_{2}(x) = 2\mathcal{A}\mathcal{B},\\
&\mathcal{P}_{3}(x) = \mathcal{A}^{2} +\mathcal{B}^{2} + \mathcal{C}.
\end{align}
where $\mathcal{A} \propto \sum_{m=1}^{L} \cos(\theta_{m})$ and $\mathcal{B} \propto \sum_{m=1}^{L} \sin(\theta_{m})$.
Thus, one reaches the relation 
\begin{equation}
\mathcal{P}^2_1(x) + \mathcal{P}^2_{2}(x)= \mathcal{P}^2_{3}(x) +\mathcal{C}. 
\end{equation} 
It results in a cone and this explains the projections shown in Fig2.(b). 

This relation also shows that $\mathcal{P}_3$ has a clear physical meaning, that is, up to a constant, it gives rise to the strength of the $U(1)$ order parameter. Since the PCA is designed to capture the largest variance in the data set, the appearance of this third eigenvalue is therefore due to that the data is generated at different temperature and the strength of $U(1)$ order varies between different temperatures. In another word, for a temperature resolved kernel PCA in which all data are generated at the same temperature, there is nearly no variation of the $U(1)$ order parameter, and therefore this eigenvalue is no longer significantly larger than the others, and there will be only two prominent components as in the linear PCA case.  

\begin{figure}[t]
\includegraphics[width=3.0 in]
{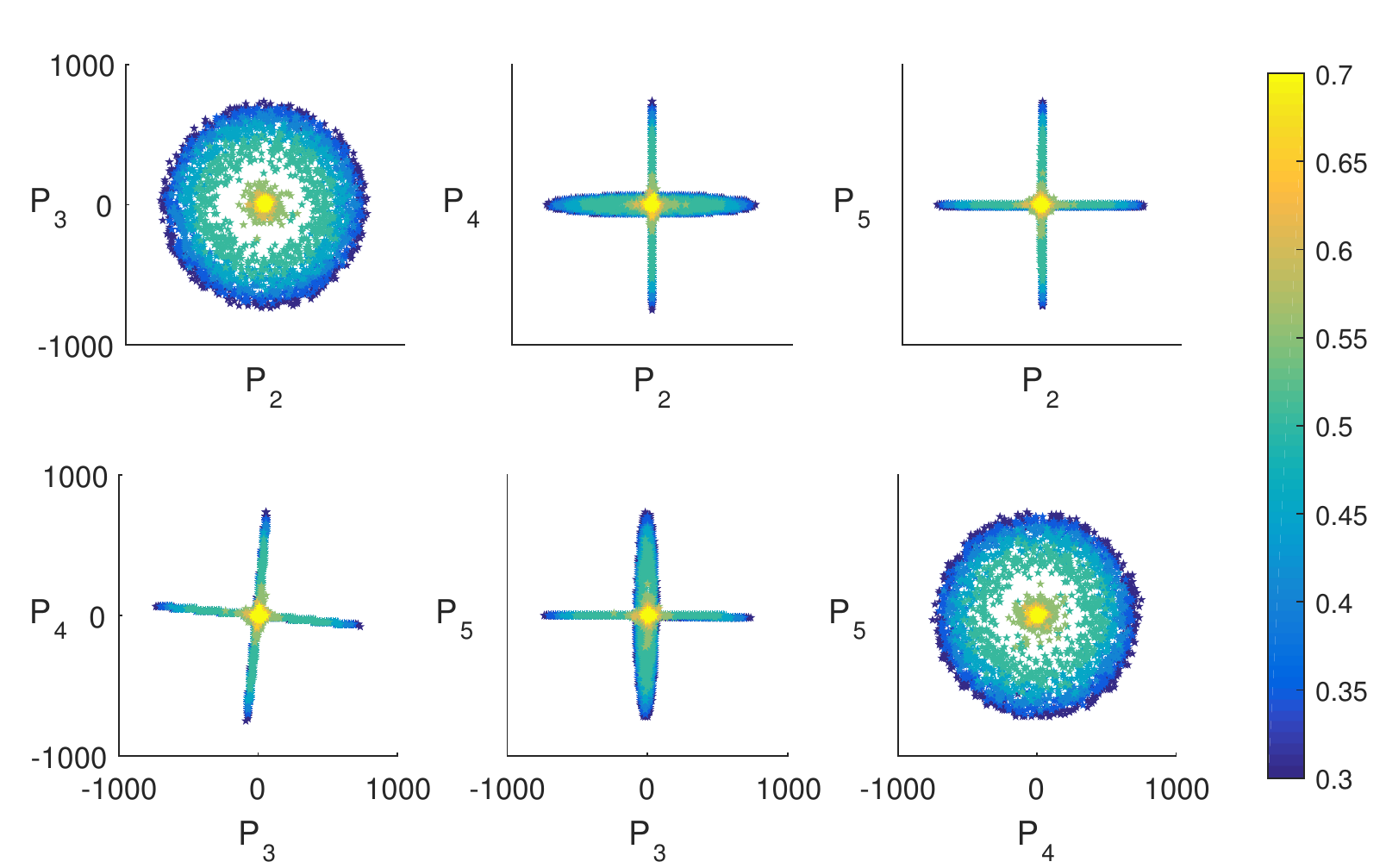}
\caption{The projections of kernel PCA onto the 2nd, 3rd, 4th and 5th components for XY model on triangular lattice. The temperature of the data set ranges from $0.3J$ to $0.7J$ , with $\Delta T = 0.05 J$. At each temperature, 1000 samples are collected.}
\label{tri_pro2345}
\end{figure}

\section{Triangular Lattice.} 

Now we move to $XY$ model with anti-ferromagnetic coupling $J>0$ in a triangular lattice. For this lattice, at temperature 
$T_{\text{c1}}=0.504J$\cite{Monte_tri, Monte_tri_uj}, a Kosterlitz-Thouless transition takes place below which there exists an algebraic $U(1)$ spin order. Below another transition temperature $T_{\text{c2}}=0.512J$\cite{Monte_tri, Monte_tri_uj}, a discrete $Z_2$ chiral order forms and the planar spins around each triangle rotate either clockwise or anti-clockwise. For triangular lattice, since these two transitions are very close, we do not have enough data between these two transition points and in practice they are regarded as a single transition. 

We apply the kernel PCA with the same quadratic kernel into data of this model generated at nine different temperatures ranging from $0.3J$ to $0.7J$ with $\Delta T = 0.05 J$. As shown in Fig. \ref{tri_eigenvalue}, we find that there are totally six eigenvalues that are significantly larger than the others, among which four (labelled as $\lambda_2,\dots,\lambda_5$) are nearly degenerate. The projection of all data $\{x_n\}$ into the subspace spanned by these four eigenvector are shown in Fig. \ref{tri_pro2345}. It looks pretty much the same as Fig. 6 of our previous papers, which shows the projection of the data into the subspace spanned by all four significantly large eigenvectors founded by linear PCA. In fact these four eigenvectors have more or less the same meaning as that founded by linear PCA. We will not repeatedly discuss them here. 

\begin{figure}[t]
\includegraphics[width=3.0 in]
{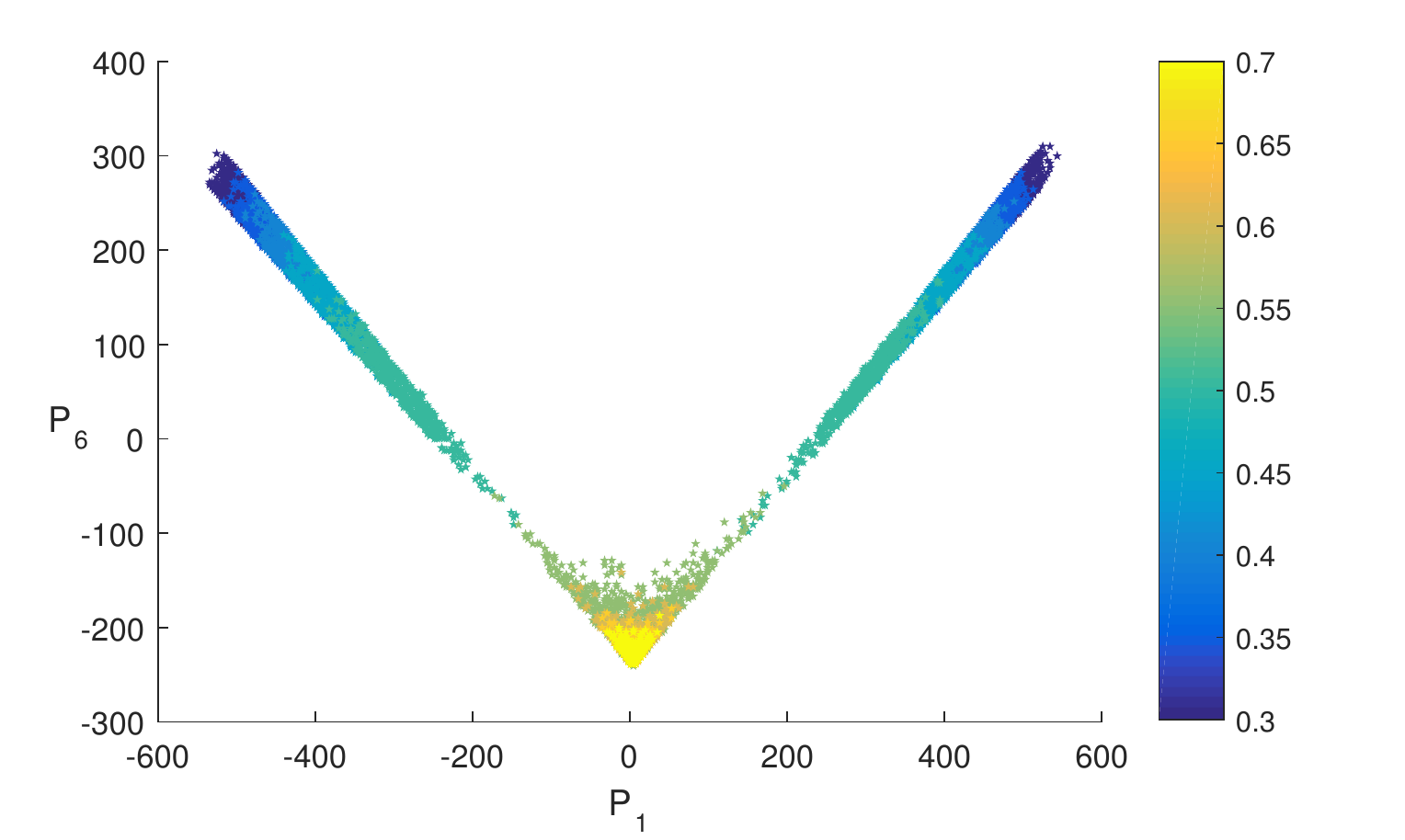}
\caption{The projections of kernel PCA  onto 1st and 6th components for XY model on triangular lattice. The temperature of the data set ranges from $0.3J$ to $0.7J$ , with $\Delta T = 0.05 J$. At each temperature, 1000 samples are collected.}
\label{tri_pro16}
\end{figure}

Here we will mainly focus on the largest eigenvalue and its corresponding eigenvector. As we show in Fig. \ref{tri_pro16}, for high temperature data, $\mathcal{P}_1$ is mainly concentrated at zero, and for low temperature data, it exhibits a symmetric double peak structure. According to what we stated as Fig. \ref{z2} in the introduction, it serves the purpose of characterizing the $Z_2$ symmetry breaking directly.  

This principal component can also be understood by a similar ``toy model". In this simple ``toy model" let us only consider three sites (labelled by A, B and C) in one unit cell. We assume that $p\%$ percent data are ordered, among which half have clock-wise chirality order and the other half have anti-clock-wise chirality order, and the rest $(1-p)\%$ precent data are completely disordered. In this case it is straightforward to show that $\mathcal{K}$ has the same structure as Eq. \ref{square_kpca}.  Because the inner-product between data with different chiral orders are zero, thus $K^{\text{low}}$ takes the form as
\begin{equation}
K^{low}=\begin{pmatrix}
K^{\uparrow} & 0\\
0 & K^{\downarrow}\\
\end{pmatrix},
\end{equation}
where 
\begin{equation}
K^{\uparrow}_{ij} =K^{\downarrow}_{ij} = \frac{L^{2}}{2}\left(1 + \cos\left(\frac{8\pi(i-j)}{pN}\right)\right),
\end{equation} 
with $L=3$.
Hence, it is straightforward to show that the largest eigenvector is 
\begin{equation}
{\bf v}_{1} = ( 1,1,...,1,-1,-1,...,-1,0,0,...0),
\end{equation}
where $ \pm 1$ correspond to the ordered data with different chirality order and $0$ is for the disordered data. According to Eq. \ref{projection},  the projection of each data onto this principal component is
\begin{align}
\mathcal{P}_{1}(x)\propto& \sum_{\theta}\Big\{\Big[\cos\left(\theta_{A} - \theta -\frac{2\pi}{3}\right)+\cos\left(\theta_{B}-\theta+\frac{2\pi}{3}\right)\nonumber\\
&+\cos\left(\theta_{C}-\theta\right)\Big]^2 -\Big[\cos\left(\theta_{A} - \theta +\frac{2\pi}{3}\right)\nonumber\\
&+\cos\left(\theta_{B}-\theta-\frac{2\pi}{3}\right)+\cos\left(\theta_{C}-\theta\right)\Big]^2\Big\} \nonumber\\
 \propto& \sin(\theta_{A}-\theta_{B})+\sin(\theta_{B}-\theta_{C})+\sin(\theta_{C}-\theta_{A})
, \label{chiral_order}
\end{align}
which is exactly the chiral order proposed for triangular lattice $XY$ model\cite{Monte_tri, Monte_tri_uj}.

Here the sixth eigenvalue in the triangular lattice $XY$ model is similar as the third eigenvalue in the square lattice $XY$ model, that represents the strength of the $U(1)$ order parameter. In Fig. \ref{tri_pro16} we show the projection of all data in the subspace spanned by the first and the sixth eigenvectors. Because the $U(1)$ transition and the $Z_2$ transition are nearly coincident with each other in this model, Fig. \ref{tri_pro16} clearly shows that, for data with $\mathcal{P}_1 \approx 0$, the $U(1)$ strength takes a minimal value; and when the $U(1)$ strength characterized by $\mathcal{P}_6$ increases toward maximum, $\mathcal{P}_1$ also shows a feature of $Z_2$ symmetry breaking. 

\begin{figure}[t]
\includegraphics[width=3.4 in]
{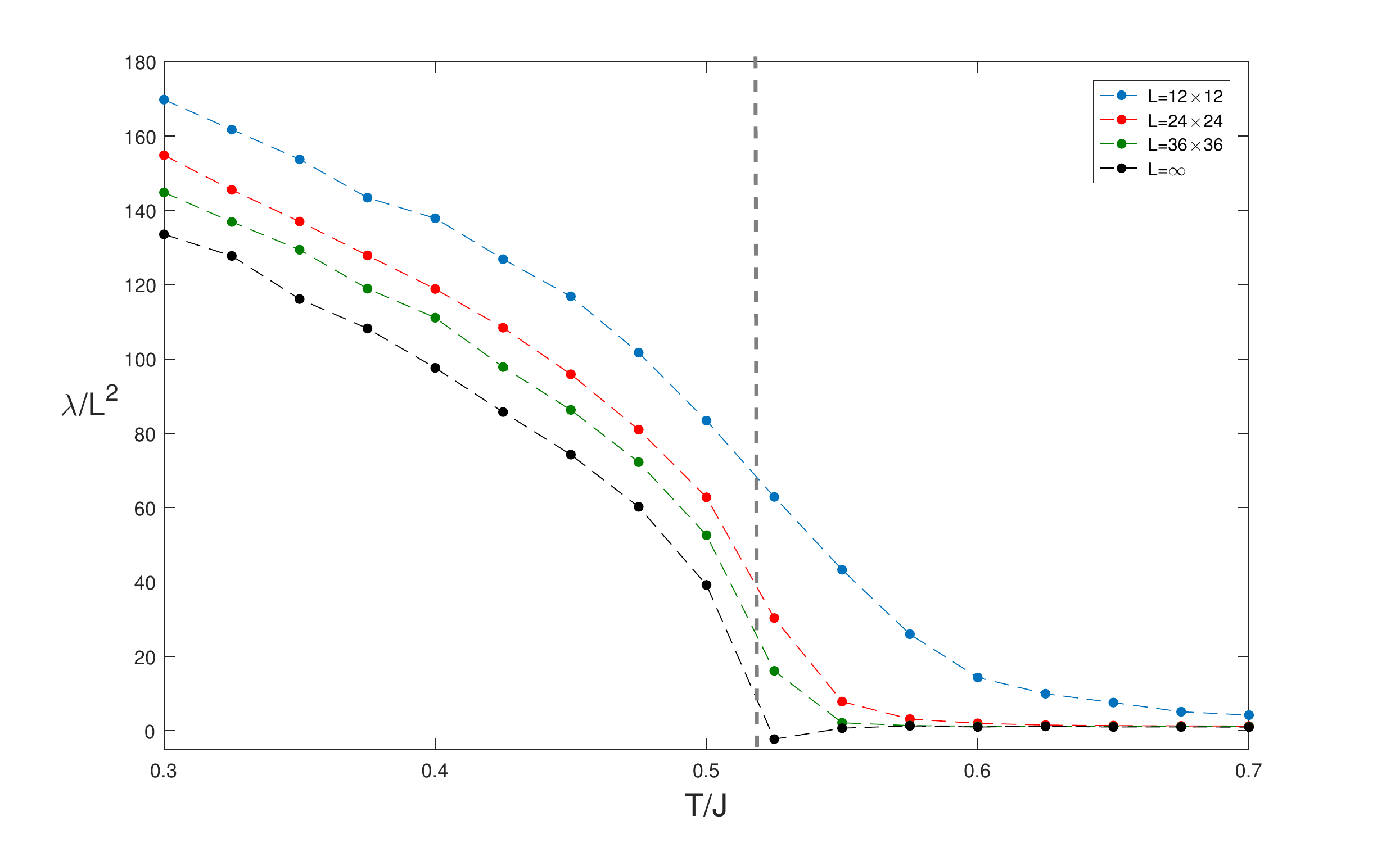}
\caption{The largest eigenvalue for temperature resolved kernel PCA .The temperature of the data set ranges from $0.3J$ to $0.7J$ , with $\Delta T = 0.025 J$. At each temperature, 1000 samples are collected. We perform a finite size scaling to get the result for $L =\infty$. }
\label{trpfss}
\end{figure}

We also consider a temperature resolved kernel PCA and we focus on the largest eigenvalue associated with this $Z_2$ chiral symmetry. At each given temperature, we consider several systems with different sizes, and then we perform a finite size scaling of this eigenvalue to infinite size limit. We then plot how the eigenvalue at infinite size limit depends on temperature. The result is shown in Fig. \ref{trpfss}, which very clearly displays a second order phase transition behavior. This is consistent with what one expects because this $Z_2$ transition fails into the Ising universality class and is characterized by a typical second-order Landau transition. 

\section{Discussion}

We can also apply the same kernel PCA scheme to the union jack lattice. The results is quite similar to triangular lattice, and we will not repeat them here. In both cases, we will find two more principal components, one of which directly reveals the $Z_2$ symmetry breaking and the other is associated with the strength of the $U(1)$ order parameters. We could also apply this method to temperature resolved case and the principal component associated with the $Z_2$ symmetry breaking exhibits clear signature of a second-order phase transition. Thus, so far we have satisfactorily addressed the $Z_2$ order and the $Z_2$ phase transition in these frustration classical spin models. 

However, there is still one more thing left. That is a common problem for both unfrustrated and frustrated classical spin models in two-dimension. That is, the $U(1)$ transition is a Kosterlitz-Thouless phase transition and can hardly be captured by these unsupervised learning methods. The deep reason is because the Kosterlitz-Thouless transition is described by vortex deconfinement and is characterized by a jump of the superfluid density\cite{XY_Tc,XY1,XY2}. Both vortex deconfinement and the supefluid density jump is related to recognization of vortex, which is a non-local topological object. Progresses have been made if vorticity or velocity fields are used as input\cite{CNN_uslearning,SqXY_PCA,KT_nn}, but so far the superfluid density jump has not been found by machine learning method using local spin configuration as input. On the other hand, a related development has been made in learning topological invariant using neural network, where winding number can be predicted accurately using local Hamiltonian as input \cite{windingnumberML}. Therefore, it conceivable that by using supervise learning method of training neural network, one can capture the Kosterlitz-Thouless transition. We leave this for future studies.

{\it Acknowledgment.} This work is supported by MOST under Grant No. 2016YFA0301600 and NSFC Grant No. 11734010.


\begin{thebibliography}{99}
\bibitem{linearPCA} C. Wang and H. Zhai, Phys. Rev. B {\bf 96}.144432 (2017).
\bibitem{Ising_PCA} L. Wang, Phys. Rev. B {\bf 94},195105 (2016).
\bibitem{Ising_nn} J. Carrasquilla and R. G.Meiko, Nat. Phy. {\bf 13}, 431 (2017).
\bibitem{Confusion} E. P. L. van Nieuwenburg, Y. H. Liu and S. D. Huber, Nat. Phy. {\bf 13}, 435 (2017).
\bibitem{Ising_BM} G. Torlai and R. G. Melko, Phys. Rev. B {\bf 94}, 165134 (2016).
\bibitem{Ising_XY_VAE} S. Wetzel, Phys. Rev. E {\bf 96}, 022140(2017).
\bibitem{Ising_SVM} P. Ponte and R. G. Melko, Phy. Rev. B {\bf 96}, 205146(2017).
\bibitem{SqXY_PCA}W. J. Hu, R. Singh and Richard Scalettar, Phys. Rev. E {\bf 95}, 062122(2017).
\bibitem{Hubbard_uslearning} K. Chng, N. Vazquez, and E. Khatami, Phys. Rev. E {\bf 97}, 013306 (2018).
\bibitem{Fermion_PCA}N. C. Costa, W. J. Hu, Z. J. Bai, Richard Scalettar and R. Singh, Phys. Rev. B {\bf 96}, 195138(2017).
\bibitem{Ising_SU(2)_nn} S. Wetzel and M. Scherzer, Phys. Rev. B {\bf 96} , 184410(2017).
\bibitem{CNN_Fermions1} K. Ch'ng, J. Carrasquilla, R. G. Melko and E. Khatami, Phys. Rev. X {\bf7}, 031038(2017).
\bibitem{CNN_Fermions2} P. Broecker, J. Carrasquilla, R. G. Melko and S. Trebst, Scientific Reports {\bf 7}, 8823(2017).
\bibitem{CNN_uslearning} P. Broecker, F.  F. Assaad and S. Trebst , arXiv:1707.00663.
\bibitem{KT_nn} M. Beach, A. Golubeva and R. G. Melko, Phys. Rev. B {\bf 97}, 045207(2018).
\bibitem{QLTML} Y. Zhang and E. Kim, Phys. Rev. Lett {\bf 118}.216401(2017).
\bibitem{Zhangyi_1}  Y. Zhang, R. G. Melko, and E. Kim, Phys. Rev. B {\bf 96}, 245119 (2017).
\bibitem{windingnumberML} P. Zhang, H. Shen and H. Zhai, Phys. Rev. Lett. {\bf 120}, 066401 (2018).
\bibitem{FXY_theory5}J. Villain, J. Phys. C 10, 1717 (1977).
\bibitem{FXY_theory6}J. Villain, J. Phys. C 10, 4793 (1977).
\bibitem{FXY_theory1}D. H. Lee, J. D. Joannopoulos, J. W. Negele, and D. P. Landau, Phys. Rev. Lett. {\bf 52}, 433 (1984).
\bibitem{FXY_theory2}S. Miyashita and H. Shiba, J. Phys. Soc. Jpn 53, 1145 (1984).
\bibitem{FXY_theory3}S. Lee and K. C. Lee, Phys. Rev. B {\bf 49}, 15184 (1994).
\bibitem{FXY_theory4}S. Korshunov, Phys. Rev. Lett. {\bf 88}, 167007 (2002).
\bibitem{FXY_theory0}M. Hasenbusch, A. Pelissetto, and E. Vicari, J. Stat. Mech.: Theory Exp. P12002 (2005) 
\bibitem{Monte_tri}T. Obuchi and H. Kawamura, J. Phys. Soc. Jpn. 81, 054003 (2012).
\bibitem{Monte_tri_uj}J. P. Lv, T. M. Garoni and Y. J. Deng, Phys. Rev. B {\bf 87}, 024108(2013).
\bibitem{XY_Tc} P. Olsson, Phys. Rev. B {\bf 52} 4526 (1995).
\bibitem{XY1} T. Ohata and D. Jasnow, Phys. Rev. B {\bf 20}.139(1978).
\bibitem{XY2} H. Weber and P. Minnhagen, Phys. Rev. B {\bf 37}.5986(1987).
\bibitem{PRML} C. M. Bishop, Pattern Recognition And Machine Learning (Springer, 2007). 



\end{thebibliography}
\end{document}